\documentstyle[11lomcon,cite,graphicx]{article}

\def\roughly#1{\mathrel{\raise.3ex\hbox
{$#1$\kern-.75em\lower1ex\hbox{$\sim$}}}}
\def\ls{\roughly<}

\begin{document}

\title{EFFECTS OF HEAVY MAJORANA NEUTRINOS AT LEPTON-PROTON COLLIDERS}

\author{ A.~Ali~\footnote{e-mail: ahmed.ali@desy.de}}

\address{Deutsches Elektronen-Synchrotron DESY, Hamburg, Germany}

\author{ A.V.~Borisov~\footnote{e-mail: borisov@ave.phys.msu.su},
D.V.~Zhuridov~\footnote{e-mail: jouridov@mail.ru}}

\address{Faculty of Physics, Moscow State University, 119992 Moscow, Russia}

\maketitle\abstracts{ We discuss the prospects of detecting the
processes $e^+p\rightarrow\bar\nu_e\ell^+\ell'^+X$ and
$\nu_ep\rightarrow e\ell^+\ell'^+X$ ($\ell,\ell'=e,\mu,\tau$)
under the conditions of the present $ep$ collider HERA and of
future colliders. These high-energy processes are assumed to be
mediated by the exchange of heavy Majorana neutrinos (HMN). We
consider two simple scenarios for the HMN mass spectrum: the effective
singlet ($m_1\ll m_2<m_3\cdots$) and the effective doublet
($m_1<m_2\ll m_3\cdots$). For the latter case, the cross section
includes information about $CP$-violating phases.}

\section*{Introduction}
At the moment there are experimental evidences for
nonzero neutrino masses \cite{Par,Giu}. The nature of neutrino
mass, whether it is Dirac or Majorana, is one of the fundamental
and still unsolved problems in particles physics. A Dirac neutrino
carries a lepton number distinguishing a particle from an
antiparticle. In contrast to that, a Majorana neutrino is
identical to its own antiparticle. The Majorana mass term in the
total Lagrangian does not conserve lepton number, but changes its
value by two units. Therefore Majorana neutrinos can lead to
various lepton number violating processes. For example, they
induce same-sign dilepton production in collisions at high
energies: $pp\rightarrow\ell^{\pm}\ell'^{\pm}X$ \cite{Ali},
$e^+p\rightarrow\tilde\nu_e\ell^+\ell'^+X$ \cite{Fla} etc.

In  theories extending the Standard Model the seesaw mechanism is
often used to provide a natural generation of small neutrino
masses (for a review, see, e.g., \cite{Lan,Kay}). Unlike the usual
way of Dirac mass generation through weak SU(2)-breaking, this
mechanism doesn't need extremely small Yukawa couplings ($\ls
10^{-12}$). For three families of leptons and $s$ right-handed
SU(2) singlets the seesaw mechanism leads to 3 light and $s$ heavy
massive Majorana neutrino states  $$
\nu_\ell=\sum\limits^3_{i=1}\tilde U_{\ell
i}\nu_i+\sum\limits^s_{j=1}U_{\ell j}N_j, $$ where $\nu_\ell$ is a
neutrino of definite flavor $(\ell=e,\ \mu,\ \tau)$, the
coefficients $\tilde U_{\ell i}$ and $U_{\ell j}$  form the
leptonic mixing matrices.

Heavy mass states give a relatively small contribution to neutrino
flavor states. Nevertheless effects of light and heavy Majorana
neutrinos (HMN) compete in lepton number violating processes,
because small values of the mixing parameters $U_{\ell j}$ for
heavy neutrinos $N_j$ may be compensated by smallness of the masses
of light neutrinos $\nu_i$.

\section*{The process $e^+p\rightarrow\bar\nu_e\ell^+\ell'^+X$}

In this report, we investigate the possibilities of observation of the
process
\begin{equation}
\label{ep}
 e^+p\rightarrow\bar\nu_e\ell^+\ell'^+X
\label{eq:process1}
\end{equation}
and its cross symmetric process $ \nu_ep\rightarrow
e\ell^+\ell'^+X $ ($X$ denotes hadron jets) under the conditions
of the present $ep$ collider HERA (DESY)\cite{Par} and of future
$ep$ colliders. We assume that these processes at high energies $$
 \sqrt{s}\gg m_W
$$ are mediated by HMN. The leading-order Feynman diagram for the
process (\ref{ep}) is shown in Fig.~1. (There is also a crossed
diagram with interchanged lepton lines.) For calculating the cross
sections, we use the leading effective vector-boson (EVB)
approximation \cite{Daw} neglecting transverse polarizations of $W$
bosons and quark mixing. For this case, cross sections for the
process and the crossed channel turn to be equal. As an
observation criteria for the process we have chosen the condition
$$ \sigma L\geq1, $$ where $\sigma$ denotes the cross section and $L$
is the integrated luminosity per year for a collider.
\begin{figure}[htb]
\vspace{-0.3cm} \centering
\includegraphics[scale=0.3]{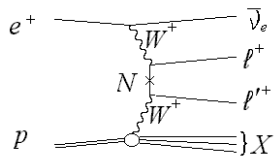}
\caption{Feynman diagram for the process
$e^+p\rightarrow\bar\nu_e\ell^+\ell'^+X$ mediated by the HMN,
$N$.} \label{Fig1}
\end{figure}

We should note that lepton-proton collisions are free of the
Standard Model background \cite{ Fla} in contrast to the
proton-proton collisions \cite{Dat}.

\subsection*{Effective Singlet Case}

At first we take the simplest pattern of the HMN mass spectrum $$
 m_{1}\ll m_{2}<m_{3}\cdots
$$ $(m_{N_i}\equiv m_i)$ assuming the condition to be held $$
\sqrt{s}\ll m_{2}. $$ The cross section
\begin{equation}\label{s}
\sigma_1=C\left(1-\frac{1}{2}\delta_{\ell\ell'}\right)|U_{\ell1}
U_{\ell'1}|^2\left(\frac{m_1}{m_W}\right)^2\int\limits_{y_0}^1\frac
{dy}{y}\int\limits_y^1\frac{dx}{x}p(x,xs)h\left(\frac{y}{x}
\right)\omega\left(\frac{ys}{m_1^2}\right)
\end{equation}
for the process is determined by the convolution of three
functions given in \cite{Ali}: $p(x,xs)$, the quark
distribution density having a fraction $x$ of the proton
momentum evaluated at the scale $Q^2=xs$, $h$, the normalized luminosity of
$W^+W^+ $ pairs in the quark-lepton system, and $\omega$, the
normalized cross section for the subprocess
$W^+W^+\rightarrow\ell^+\ell'^+$. Here, $y_0=4m_W^2/s$ and the
characteristic constant $C$ has the value
\begin{equation}\label{a}
C={G_F^4m_W^6}/({8\pi^5})=0.80~{\rm fb}.
\end{equation}

In the numerical calculation we have used the set of parton
distributions CTEQ6 \cite{Pum}. Using the bounds on the mixing parameters
$U_{\ell N}$ from precision electroweak data \cite{Nar}
\begin{equation}\label{sum}
\sum\limits_N|U_{eN}|^2<6.6\times 10^{-3}, \sum\limits_N|U_{\mu
N}|^2<6.0\times 10^{-3}, \sum\limits_N|U_{\tau
N}|_{eff}^2<3.1\times 10^{-3}
\end{equation}
and the constraint from the neutrinoless double beta decay
\cite{Bel}
$$
 \left|\sum\limits_NU_{eN}^2m_N^{-1}\right|<5\times 10^{-5}~{\rm TeV}^{-1}
$$
(the sum is over the heavy neutrinos), we find that the process
is practically unobservable at HERA even with a very optimistic
luminosity ($\sqrt{s}=318~{\rm GeV},\; L=1~{\rm fb}^{-1}$) and
also at the projected supercollider VLHC (see, e.g., \cite{Alm})
($\sqrt{s}=6320~{\rm GeV}$, $L=1.4~{\rm fb}^{-1}$). For example,
for $m_1  \sim 1~{\rm TeV}$, we get $\sigma L \sim 10^{ -
10}~(10^{ - 3} )$ for HERA (VLHC). For a possible detection of the
process, the luminosity and/or the energy of the $ep$-collider
should be substantially increased. Taking for example the
luminosity $L=100~{\rm fb}^{-1}$, the observation of the most
probable events ($\mu\tau$ and $\mu\mu$) is possible if $\sqrt{s}>
23~{\rm TeV}$. For $\sqrt{s}=25~{\rm TeV}$ such a collider will be
sensitive to a range of neutrino masses about 1--3 TeV.
\subsection*{Effective Doublet Case}
We consider also the neutrino mass spectrum of the
effective doublet type $$
 m_1<m_2\ll m_3\cdots
$$ with the bound on energy $\sqrt{s}\ll m_3$. In this scenario,
the cross section for the process (\ref{eq:process1})
\begin{equation}
\label{c2} \sigma _2  = \frac{C}{2}\int\limits_{y_0 }^1
{\frac{{dy}}{y}} \int\limits_y^1 {\frac{{dx}}{x}} p(x,xs)h\left(
{\frac{y}{x}} \right)W\left( {\frac{{ys}}{{m_1^2
}},\frac{{ys}}{{m_2^2 }}} \right)
\end{equation}
includes the normalized cross section for the subprocess
\begin{equation}
\label{W} W(t_1 ,t_2 ) = m_W^{-2}\left[ {\rho _1^2 m_1^2
\omega (t_1 ) + 2c\rho _1 \rho _2 m_1 m_2 \Omega (t_1 ,t_2 ) +
\rho _2^2 m_2^2 \omega (t_2 )} \right],
\end{equation}
which contains the individual contributions of the neutrinos $N_1$ and
$N_2$, $\omega (t_i )$,  and the interference of the two, $\Omega(t_1,t_2)$,
where
 $$
\Omega(t_1,t_2)=2-\frac{1}{t_1+t_2+t_1t_2}\left[\frac{t_2(t_1^2
-2t_1t_2-2t_2)}{t_1(t_1-t_2)}\ln(1+t_1) + (t_1 \leftrightarrow
t_2)\right];$$ $$\omega(t)=\lim\limits_{t'\rightarrow
t}\Omega(t,t')=2 + \frac{1}{{1 + t}} - \frac{{2(3 + 2t)}}{{t(2 +
t)}}\ln (1 + t). $$ The mixing parameters for different
$\ell\ell'$ channels of the process are $$
 \rho_i=\sqrt{2-\delta_{\ell\ell'}}\left|U_{\ell i}U_{\ell'i}\right|,\quad
 c=\cos\delta_{\ell\ell'}
$$ with $$
 \delta_{\ell\ell'}=\phi_1-\phi_2\in[0,2\pi),\quad\phi_i=\arg\left(U_{\ell
i}U_{\ell'i}\right). $$ The phases $\delta_{\ell\ell'}$ carry
information about CP-violation.
\begin{figure}[htb]
\vspace{-0.5cm}
\centering
\includegraphics[scale=0.55]{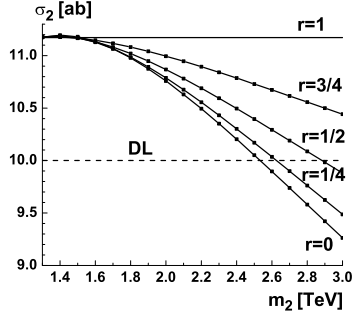}
\vspace{-0.5cm}
\caption{The dependence of $\sigma_2$ (in attobarn) on $m_2$
(in TeV) plotted for
$r=0,~1/4,~1/2,~3/4,~1$ with $\sqrt{s}=25$ TeV, $m_1=1.3~{\rm
TeV}$ and $c=1$. The horizontal line DL is the discovery limit.}
\label{Fig2}
\end{figure}
\vspace{-0.2cm}
We assume the saturation of the upper bound $B=6.0\times 10^{-3}$
in the second sum in (\ref{sum}) only by the first two terms, i.e.,
$|U_{\mu 1} |^2  = rB,\;|U_{\mu 2} |^2  = (1 - r)B$ with $r\in
[0,1]$. Then for the most probable $\mu\mu$ channel we obtain
\begin{equation}
\label{c2x} \sigma _2  = A(r^2 f_1  + 2cr\bar rF_{12}  + \bar r^2
f_2 ),\quad \bar r = 1 - r,
\end{equation}
where $f_i=f(s,m_i)$ and $F_{ij}=F(s,m_i,m_j)$ are expressed
through obvious convolutions of the functions $\omega$ and $\Omega$
with $h$ and $p$, respectively (see Eq. (\ref{c2})), the
constant $A$ has the value $A = 1.4\times10^{-5}~{\rm fb}$. For $r=1~(r=0)$, only
a single neutrino $N_1$ ($N_2$) contributes to the cross section
which is reduced to the form given in  Eq. (\ref{s}). Generalization to
the case of $n$ neutrinos is straightforward.

In our calculations, we have chosen the following values for the
parameters: $\sqrt{s}=25$ TeV, $m_1=1.3$ TeV, $c=1$. The cross
section (\ref{c2x}) as a function of $m_2$ for various fixed
values of $r$ is shown in Fig.~2. For the almost degenerate doublet
($m_1\simeq m_2$) case and/or for the case of small mixing with $N_2$
($r\simeq 1$), the cross section $\sigma_2$ is close to
$\sigma_1$, the cross-section for the effective singlet
case. But we should note that for
the case of destructive interference of the two almost degenerate
massive states (e.g., for $m_2\simeq m_1$, $r=1/2,~c=-1$), the
cross section is vanishingly small.

\section*{Acknowledgments}
\vspace{-0.2cm}
We thank Dmitri Peregoudov for help in numerical
calculations.

\section*{References}


\begin{thebibliography}{99}

\bibitem{Par} Particle Data Group Collab.: K.~Hagiwara et al.,
 {\it Phys. Rev.} {\bf D66}, 010001 (2002).

\bibitem{Giu} C.~Giunti, E-print Archive: hep-ph/0310238;\\
 A.Yu.~Smirnov, E-print Archive: hep-ph/0311259.

\bibitem{Ali} A.~Ali, A.V.~Borisov, N.B.~Zamorin, in {\it ``Frontiers of Particle Physics"}
(Proceedings of the 10th Lomonosov Conference on Elementary
Particle Physics), ed. by A.~Studenikin (World Scientific,
Singapore,~2003) p.~74;\\
{\it Eur.\ Phys.\ J.}  {\bf C21}, 123 (2001)
  [hep-ph/0104123].

\bibitem{Fla} M.~Flanz, W.~Rodejohann, K.~Zuber, {\it Phys. Lett.} {\bf B473}, 324 (2000);\\
W.~Rodejohann, K.~Zuber, {\it Phys. Rev.} {\bf D62}, 094017
(2000).

\bibitem{Lan} P.~Langacker, {\it Nucl. Phys. B (Proc.
Suppl.)} {\bf 100}, 383 (2001).

\bibitem{Kay} B.~Kayser, E-print Archive: hep-ph/0211134.



\bibitem{Daw} S.~Dawson, {\it Nucl. Phys.} {\bf B249}, 42 (1985);\\
I.~Kuss, H.~Spiesberger, {\it Phys. Rev.} {\bf D53}, 6078 (1996).

\bibitem{Dat} A.~Datta, M.~Guchait, D.P.~Roy, {\it Phys. Rev.}
{\bf D47}, 961 (1993).

\bibitem{Pum} J.~Pumplin, D.R.~Stump, J.~Huston, H.L.~Lai, P.~Nadolsky,
W.K.~Tung, E-print Archive: hep-ph/0201195.

\bibitem{Nar} E.~Nardi, E.~Roulet, D.~Tommasini, {\it Phys. Lett.} {\bf B344}, 225
(1995).

\bibitem{Bel} G.~Belanger, F.~Boudjema, D.~London, H.~Nadeau, {\it Phys. Rev.}
{\bf D53}, 6292 (1996).

\bibitem{Alm} M.~Blaskiewicz et al., Fermilab Report TM-2158, 29 June 2001;
\\F.M.L.~de Almeida~Jr., Y.A.~Coutinho, J.A.~Martins Sim\~oes,
M.A.B.~do Vale, {\it Phys. Rev.} {\bf D65}, 115010 (2002).

\end{thebibliography}
\end{document}